\documentclass[electronics,review,accept,moreauthors,pdftex]{mdpi} 

\usepackage{subfigure} 
\usepackage{xcolor}
\newcolumntype{C}[1]{>{\PreserveBackslash\centering}p{#1}}
\newcolumntype{R}[1]{>{\PreserveBackslash\raggedleft}p{#1}}
\newcolumntype{L}[1]{>{\PreserveBackslash\raggedright}p{#1}}

\firstpage{1}
\pubvolume{1}
\issuenum{1}
\articlenumber{0}
\doinum{}
\pubyear{2022}
\copyrightyear{2022}
\datereceived{} 
\dateaccepted{} 
\datepublished{10.3390/electronics11030410} 
\hreflink{} 

\Title{3-D Metamaterials:  Trends on Applied Designs, Computational Methods and Fabrication Techniques}

\TitleCitation{3-D Metamaterials:  Trends on Applied Designs, Computational Methods and Fabrication Techniques}

\Author{Antonio Alex-Amor 
 $^{1,}$*, Ángel Palomares-Caballero $^{2}$ and Carlos Molero $^{2}$}

\AuthorNames{Antonio Alex-Amor, Ángel Palomares-Caballero and Carlos Molero}

\AuthorCitation{Alex-Amor, A.; Palomares-Caballero, Á.; Molero, C.}

\address{%
$^{1}$ \quad Information Technology Department, Universidad CEU San Pablo, 28668 Madrid, Spain \\ 
$^{2}$ \quad Department of Signal Theory, Telematics and Communications, University of Granada, 18071 Granada, Spain;  angelpc@ugr.es (Á.P.-C.); cmoleroj@ugr.es (C.M.)
}

\corres{Correspondence: antonio.alexamor@ceu.es}

\abstract{Metamaterials are artificially engineered devices that go beyond the properties of conventional materials in nature. Metamaterials allow for the creation of negative refractive indexes; light trapping with epsilon-near-zero compounds; bandgap selection; superconductivity phenomena; non-Hermitian responses; and more generally, manipulation of the propagation of electromagnetic and acoustic waves. In the past, low computational resources and the lack of proper manufacturing techniques have limited  attention towards 1-D and 2-D metamaterials. However, the true potential of metamaterials is ultimately reached in 3-D configurations, when the degrees of freedom associated with the propagating direction are fully exploited in design. This is expected to lead to a new era in the field of metamaterials, from which future high-speed and low-latency communication networks can benefit. Here, a comprehensive overview of the past, present, and future trends related to 3-D metamaterial devices is presented, focusing on efficient computational methods, innovative designs, and functional manufacturing techniques.
}


\begin{document}

\section{Introduction} \label{sec1}

\begin{figure}[H]
    \hspace*{1.9cm}
    \subfigure{\includegraphics[width= 0.65\textwidth]{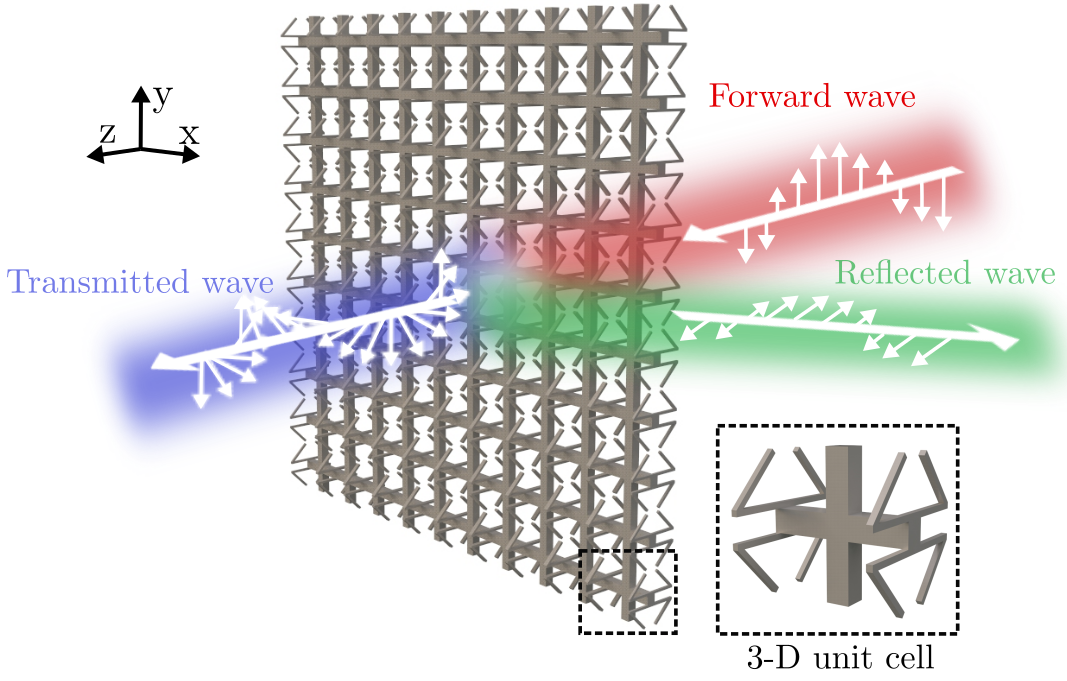}
	}
    \caption{A conceptual illustration of a 3-D metamaterial controlling the flow of~light. } 
	\label{conceptual_example}
\end{figure}

The appearance of metamaterials has supposed one of the biggest revolutions in the field of electromagnetics in this century~\cite{metamaterial_pendry}. Metamaterials are human-made composite structures that allow for tailoring the propagation of electromagnetic and acoustic waves in media. These artificial structures go beyond the properties of their constitutive materials; namely, the~geometrical disposal of their elements (frequently named ``meta-atoms'') determines to a great extent their mechanical or electrical response, similar to crystals and protein chains. The~most outstanding point about metamaterials is that they can be engineered to exhibit unusual properties rarely found in nature~\cite{metamaterial_book, metamaterials_review}, i.e.,~artificial magnetism (magnetism without inherent magnetic materials), negative-refractive indexes from positive-index materials, invisibility cloaking (``invisible'' materials that do not interact with light), non-reciprocal phenomena, and chiral responses, such as~the one schematized in Figure~\ref{conceptual_example}.

Up until now, communities of scientists and engineers have put their efforts into one-dimensional (1-D) and two-dimensional (2-D) metamaterials~\cite{metasurfaces_review, metasurfaces_review2}, leaving three-dimensional (3-D) configurations  to the side. The~reason for this is the high design complexity involved in 3-D metamaterials compared to its 1-D and 2-D counterparts. Nonetheless, the~increased complexity is worth the price to pay.

\newpage
Three-dimensional metamaterials present advantages compared with one- and two-dimensional metasurfaces~\cite{metamateriales3D_review}. In general, these advantages are attributed to the fact that a 3-D structure allows the degrees of freedom in both the transversal and longitudinal directions to be exploited in design. In other words, there are more geometrical parameters to tune the behavior of the device. Thus, 3-D metamaterials lead to \emph{advanced functionalities} that are difficult to achieve with 1-D and 2-D configurations, from which future communication systems and smart radio environments can benefit~\cite{gradoni2021smart}. An example of advanced functionality is the independent control of two orthogonal polarization states, which permits the same structure to act simultaneously as a beam scanner (splitter/steerer) and  wave absorber~\cite{Molero2021}. Moreover, the efficiency can be improved in 3-D structures by avoiding the use of dielectric materials and by directly considering fully metallic implementations~\cite{Molero2020}. This is not possible in planar structures manufactured with traditional printed circuit board (PCB) techniques, where the metasurface is attached to a dielectric substrate. In addition, 3-D structures can be easily scaled to other frequency ranges as long as fully metallic designs are considered.

\begin{figure}[H]
    \subfigure[]{\includegraphics[width= 0.35\textwidth]{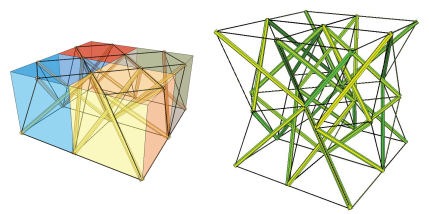}
	} 
	\subfigure[]{\hspace*{1.6cm}\includegraphics[width= 0.35\textwidth]{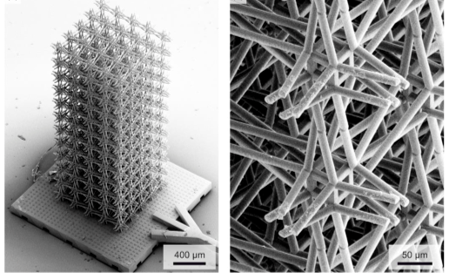}
	}\\
	\subfigure[]{\includegraphics[width= 0.37\textwidth]{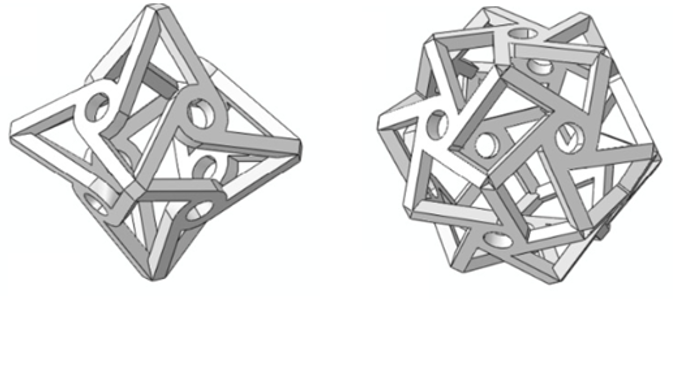}
	}
	\hspace*{1.9cm}
	\subfigure[]{\includegraphics[width= 0.23\textwidth]{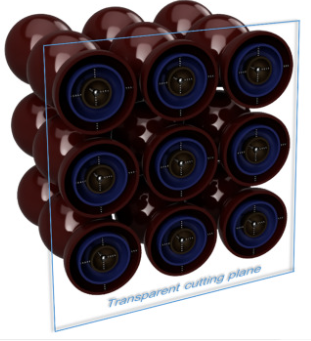}
	}\\
	\hspace*{0.7cm}\subfigure[]{\includegraphics[width= 0.29\textwidth]{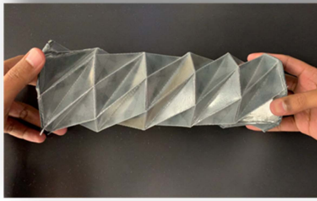}
	}
	\subfigure[]{\hspace*{1.3cm}\includegraphics[width= 0.43\textwidth]{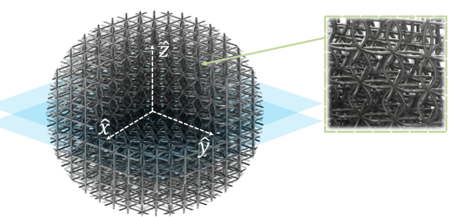}
	}
    \caption{Examples of 3-D metamaterials. (\textbf{a}--\textbf{c}) Mechanical metamaterials 
~\cite{cc1, cc2, cc4}. (\textbf{d}) Acoustic metamaterial~\cite{cc5}. (\textbf{e}) Antenna~\cite{cc3}. (\textbf{f}) Electromagnetic metamaterial~\cite{cc7}.} 
	\label{collage}
\end{figure}

Naturally, these advantages come at a price. Three-dimensional metamaterials are bulkier than planar structures, and for some purposes, this can be prohibitive. They also demand greater computational resources and are significantly more complex to manufacture, which have limited their implementation in the past. This trend has changed in recent years, thanks to the massive technological progress in computer electronics (increased computational power) and the development of 3-D printing techniques. This can be appreciated in Figure~\ref{collage}, where metamaterials of different fields (mechanics, acoustics, and electromagnetics) are~illustrated.

In this paper, we present an overview of the most outstanding recent developments related to 3-D metamaterials. The~document is organized according to the following contents. In~Section~\ref{sec2}, a~review of the most relevant computational techniques applied to 3-D metamaterials is presented. In~Section~\ref{sec3}, we discuss on some relevant works that have exploited the advantages of 3-D metastructures for the design of functional electromagnetic devices. In~Section~\ref{sec4}, we review the fundamental fabrication techniques for 3-D prototyping.  In~Section~\ref{sec5}, we discuss on some future challenges and promising approaches related to 3-D design.  Finally, conclusions are drawn in Section~\ref{sec6}.
\vspace{-12pt}
\unskip

\section{Computational~Methods}\label{sec2}
In a world where mass device prototyping takes a long period of time and costs thousands or million dollars, computational methods are fundamental to predicting the electromagnetic behavior of devices and, therefore, to reduce costs in manufacturing. Regretfully, there is no perfect method that meets all of the simulation requirements (accuracy, fast computation, robustness, flexibility, etc.) and adapts to all scenarios in an efficient manner. Each method possesses \emph{strengths and weaknesses
} that make it appropriate for {specific domains and unsuitable for others~\cite{review_computational}}. Thus, it is a common situation to find hybrids of different methods to improve computation efficiency. This section reviews the most relevant computational electromagnetic methods for 3-D structure design and highlights the domain of interest where these methods can be smartly~applied.

\subsection{Differential-Form~Methods}

Maxwell's equations can be understood in either integral or differential forms~\cite{fullwave_methods}. The~latter seeks to represent electromagnetic field propagation as a set of differential equations. Logically, analytical solutions can only be found in very restricted and simple scenarios. Thus, numerical approaches are required to give approximate solutions to the configurations under study. Numerical solutions based on differential-form methods are often referred to as full-wave approaches, with the two most popular tools being the finite difference method and the finite element~method.

The simplest full-wave technique used to solve the complete set of Maxwell's equations is the \emph{finite difference method} \cite{FDTD_general, fdtd_matlabbook}. The~original work developed by Yee in 1966~\cite{FDTD_Yee} sought to transform the original Maxwell's equations into difference equations to solve a variety of 3-D electrogmanetic problems. This would be a finite-difference time-domain (FDTD) scheme. Nonetheless, finite differences can be directly applied in the frequency domain when time-harmonic fields are considered. This would be a finite-difference frequency-domain (FDFD) scheme~\cite{FDFD}. Comparing both, FDTD is normally preferable to analyze transient and nonlinear phenomena while FDFD is a better option for electrically large structures. As~an example, FDTD schemes are implemented in~\cite{FDTD_cavity, FDTD_cavity2, FDTD_woodpile} for the analysis of 3-D (nano) cavities and woodpile structures, respectively. FDFD techniques have also been implemented to model the scattering phenomena in isotropic~\cite{FDFD_iso} and anisotropic~\cite{FDFD_aniso} 3-D objects, and the dispersion properties of open nanophotonic resonators~\cite{FDFD_nano}. 

One of the most widely implemented technique in commercial packages is the \emph{finite element method} (FEM). This is essentially due to its facility to deal with non-trivial scenarios in a more efficient way than a traditional finite difference scheme in~exchange for a greater complexity in its mathematical implementation~\cite{fem_book}. Similar to FDTD and FDFD, the~whole space in FEM is discretized into small pieces, in~this case, called finite elements. In~a Galerkin approach of the FEM, the~residual of the so-called weak form is forced to be orthogonal to some predefined test functions and the fields are expanded as a set of basic functions. This leads to a finite system of linear equations that can be solved with direct or iterative methods. Some examples of in-house FEM formulations applied to electromagnetic 3-D design can be found in the literature. In~\cite{fem3d_cloak}, a~FEM was developed for the design of efficient cloaks applied to spheroidal metallo-dielectric objects. Another example can be found in~\cite{FEM3D_crystals}, where a FEM was applied to compute EM fields in 3-D nano-photonic crystals and waveguides. However, as~the FEM is such a generalist method, the~common scenario is to perform FEM computations through commercial simulators as CST, COMSOL, or Ansys HFSS. This is the case in~\cite{FEM3D_chiral}, where the proposed analytical model for chiral 3-D elastic metamaterials is checked with FEM calculations performed with commercial simulator COMSOL Multiphysics~\cite{comsol}.

\subsection{Integral-Equation~Methods}

Different from finite difference and FEM schemes, integral-equation methods incorporate radiation conditions in a natural form, so there is no need to include absorbing boundary conditions to restrict the simulation domain~\cite{integral_article}.  In~addition, integral equations allow for a reduction in the dimension of the problem by one by only setting unknowns on the boundaries instead of discretizing the entire volume~\cite{integral_equations}. These motives have propelled the use of integral-equation methods in branches of electromagnetism such as antenna theory or microwave circuits~\cite{book_integral}.

The most famous integral-equation method is known as the \emph{method of moments} (MoM), occasionally named the boundary element method in other fields of physics~\cite{book_mom}. In~the MoM, the~unknown function  $f$ to be determined is expanded inside the integral as a set of known basis functions $b_k$  with unknown coefficients $a_k$; namely, $f(\mathbf{r},t) \approx \sum_k a_k b_k(\mathbf{r},t)$. This eventually leads to a linear system of equations with unknown coefficients $a_k$ that can be solved after considering the appropriate boundary conditions. MoM has been traditionally applied in the study of {wire and planar antennas~\cite{book_mom, mom_wire, mom_metasurface}} as~well as in single-layer reflectarray/transmitarray configurations~\cite{mom_reflectarray}. These works were then extended to include planar multilayered (2.5-D) devices~\cite{MoM_Florencio1, MoM_Florencio2, MoM_Florencio3, MoM_Corcoles}. However, it is not so common to find actual 3-D implementations with the MoM, which is~a difference from full-wave numerical methods. This is mainly due to the greater difficulty in defining proper basis functions for complex 3-D geometries and to the substantial increase in time complexity compared with 1-D and 2-D structures. Nonetheless, some interesting examples can be found in the literature. For~instance, MoM formulations are developed in~\cite{mario3D_IE} (combined with a plane-wave expansion) for an analysis of 3-D metallic wire media and in~\cite{craeye3D_IE} to compute the eigenstates of periodic metamaterials, with~application to 3-D arrays of~spheres.  

\emph{Nyström method} is an alternative to method of moments. In the~Nyström (quadrature) method, integrals are directly replaced by weighted sums~\cite{nystrom_book}. It is generally more computationally costly compared with MoM but can also offer more accurate results as higher-order quadrature rules can be easily implemented.  The Nyström method is particularly convenient compared with MoM when a mixture of several
media of different characteristics has to be considered in the computation. An~example of its use applied to the study of scattering phenomena in 3-D objects and composite structures can be found in~\cite{nystrom3D1, nystrom3D2}.

The \emph{fast multipole method} (FMM) is an integral-equation method extensively used for the resolution of gravitational and electromagnetic problems~\cite{fastmultipole}. The~essential idea behind the FMM is to locally approximate a simple radiating point by superposing a finite number of plane/spherical waves in the direction of propagation; namely, expanding the system Green's function using multipole series terms. Different aspects related to the numerical implementation are extensively detailed in~\cite{fastmultipole2}. Since the FMM is a quite efficient computational technique to perform matrix–vector products, it is commonly found in combination with basic MoM methods to speed up the resolution of the linear system. An~example can be found in~\cite{fastmultipole3D}, where a hybrid FMM--MoM method is employed to analyze the scattering of electromagnetic waves in 3-D nanoparticle~systems.

The \emph{Wiener--Hopf method} is  a powerful technique for solving certain types of integral equations. By~making use of the Laplace transform ($Z$ transform in discrete implementations), it is possible to transform the original expression into a functional equation that is suitably defined in the complex space~\cite{wienerhopf}. In~the context of electromagnetics, its use is convenient in periodic problems, where the periodicity is considered a space sampling. Concretely, the Wiener--Hopf method has proven to be particularly insightful and efficient in truncated periodic configurations, such as~the semi-infinite 3-D  plasmonic media presented in~\cite{wienerhopf3D}. Thanks to the use of this method, edge diffraction terms and bound surface waves launched by the edge condition (terms that are not excited in infinite structures) have been clearly identified~\cite{wienerhopf_camacho}.

Worthy of special mention is \emph{Ewald's method}, traditionally used for the efficient computation of free-space periodic Green's functions~\cite{review_ewald}. The~underlying idea of Ewald's summation method is to rewrite a generic periodic Green's function as a sum of two quickly-convergent series: the first sum (actual space) exponentially decreases by itself, and the second sum (Fourier-reciprocal space)  quickly decays for large arguments. Ewald's method has been typically applied to compute the dispersion properties in 3-D periodic arrays of electric and magnetic dipoles~\cite{ewald3D, ewald3D2} and in 3-D  skewed lattices~\cite{ewald3D_mosig}.

\subsection{Modal~Analysis}

In modal analysis techniques, the~fields are expanded as a series of orthogonal eigenfunctions weighted by complex coefficients. The~amplitude of these coefficients are fixed by the boundary conditions and the geometry of the considered object. As~a difference with purely numerical approaches, modal techniques are direct and bring physical insight to wave propagation and scattering phenomena while maintaining fast computation times~\cite{modal_wexler}.

Modal techniques are normally applied in closed structures where the fields are confined, such as waveguides and coaxial lines~\cite{collin}. In~fact, the~analysis of waveguide junctions and discontinuities is one of the most characteristic applications of modal analysis, which raised a lot of interest in the engineering community for the design of filters, diplexers, orthomode transducers, and~other interesting microwave devices. This is performed by imposing the continuity conditions of the tangential fields in~a process that is normally referred to as \emph{mode matching} (MM) \cite{modematching}. 

In addition, modal techniques are also intensively applied for the analysis and design of metamaterials. Plane waves are solutions to Helmholtz wave equation and form a proper basis to represent EM fields in periodic structures~\cite{modal_planewaves}. \emph{Plane-wave expansions} (also known as Floquet--Bloch expansions) are modal representations of the fields in the region of interest. Thanks to the Floquet--Bloch theorem, the~whole computation domain can be reduced to a single unit cell in periodic configurations and,~therefore, the plane-wave expansion, notably reducing the computational requirements of the~simulation. 

Multiple examples of modal expansions applied to metamaterials and antennas can be found in the literature. In~\cite{modal_spheres, modal_spheres2}, a~modal analysis was carried out to compute the dispersion properties of 3-D periodic arrangements of spheres, with~potential application to complex nanoparticle systems. A~MM technique was formulated in~\cite{jaime_modematching} for the analysis of bi-periodic cylindrical structures and wire media. In~\cite{modematching_fdfd}, a~hybrid MM-FDFD was developed for the analysis of electromagnetic bandgap (EBG) structures and substrate-integrated-waveguide (SIW) antennas. In~\cite{epsteinMM}, dual-polarized fully metallic metagratings constituted by short-circuited periodic waveguides were analyzed with a MM technique. Periodic arrays of helical structures forming chiral metamaterials were studied in~\cite{chiral_helix} by means of modal analysis techniques.  In~addition, efficient semi-analytical formulations based on the mode-matching method and Floquet modal expansions were developed recently for the analysis of metasurfaces and transmission lines involving higher symmetries (glide and \mbox{twist) \cite{mm_antonio1, mm_antonio2, mm_oskar, mm_sipus}.} These formulations have potential to be adapted to mimic the interesting properties associated to higher symmetries (low dispersion, wideband, and higher refractive index) in more complex 3-D~structures.

\subsection{Circuit~Models}
Circuit models are insightful tools that allow for a simple representation of the underlying physical reality. Interestingly, they are noticeably more computationally efficient compared with the aforementioned methods, most of the time, at~the cost of a lower accuracy in the computation. These features make circuit models a suitable option for the modeling of microwave devices, frequency selective surfaces, and metamaterials~\cite{book_marcuvitz, mesa_review, costa_review}.

There are different methods to obtain a \emph{circuit model} in periodic configurations: (i) heuristic rationale, namely, simple circuits based on experimental knowledge (capacitive/inductive coupling, losses modeled as resistances, etc.) that provide physical insight but work for very particular structures, usually in a limited frequency range~\cite{heuristics1, heuristics2}; (ii) synthesis methods, that is,  more rigorous (and complex) formulations that give accurate results but require prior full-wave simulations~\cite{sintesis_page, sintesis_costa, sintesis_jaime}; and (iii) analytical methods, namely, rigorous circuit approaches, {based on Floquet modal expansions~\cite{floquet_raul, floquet_frezza, floquet_hum} or first-principle computations (i.e., integral equations) \cite{circuit_integral1, circuit_integral2}, that provide accurate results and are independent from commercial full-wave simulators but are often restricted to canonical geometries.}

Recent advances in circuit modeling have considered stacks of periodic 2-D \linebreak arrays~\cite{sintesis_page,  carlos25D} and the potentials of these methods are somehow broader than expected~\cite{antonio25D}. Nonetheless, stacks of 2-D arrays do not really exploit the design the degrees of freedom associated with the longitudinal (propagating) direction. For~this reason, they are often referred to as 2.5-D structures. In~fact, there is still a lack of studies in the literature on general circuit approaches that can model 3-D configurations. Only a few  selected examples can be found. In~\cite{grbic3D_circuit}, the~proposed isotropic 3-D negative-refractive-index medium was modeled  by means of a circuit model consisting in transmission lines loaded with reactive elements. In~\cite{caloz3D_circuit, russer3D_circuit}, a~circuit model based on the transmission-line matrix  scheme {(TLM, see~\cite{TLM_3D})} was proposed for 3-D composite right-left-handed isotropic~metamaterials. 

{The design of 3-D square waveguide cells for implementing full-metal devices was recently investigated by the authors. In~\cite{molero_circuit3D}, a~quasi-analytical circuit approach that combines analysis and synthesis procedures was developed for the design of full-metal polarizers. In~\cite{circuit_carlosUGR}, a~periodic 3-D metamaterial formed by laterally perforated square waveguides that exploit phase resonances was analyzed  by means of a heuristic circuit approach. Finally, a~heuristic circuit (assisted by full-wave computations in commercial software CST) was presented in~\cite{molero3D_2022} for the simplified design of complex broadband 3D-printed polarizers based on electro-magnetic elements. }

\subsection{Other~Methods}
Although the most popular and generalist methods have already been discussed, there are many other solutions that are particularly efficient when applied to some particular configurations and structures. This would be the case of the homogenization, ray-tracing, and transfer-matrix~methods.

\emph{Homogenization theory} seeks to model the electromagnetic behavior of complex media through their effective (average) constitutive parameters. Homogenization methods based on first-principle computations and the use of integral equations have been particularly useful for the analysis of {3-D wire media~\cite{blanchard3D, mario3D, mario_wireboundary, hom3D_silveirinha}}, epsilon-near-zero materials~\cite{marioENZ},  {3-D resonators~\cite{hom_resonator}}, and~3-D non-magnetic configurations~\cite{mario3D_hom}. A~different example of homogenization model can be found in~\cite{guido_hom}, where the effective parameters of a  glide-symmetric metamaterial were derived with a mode-matching~formulation.

\emph{Ray theory} describes light propagation and, more generally, wave propagation in terms of rays, a~geometrical simplification to represent light paths~\cite{ray_general}. Ray-tracing methods are efficient tools used to estimate propagation features in very large domains and multi-scale problems, where FEM or finite difference methods require huge computational resources.  In~contrast, ray theory is valid as long as the wavelength is small compared with the dimensions of the considered objects. In~electromagnetics, ray-tracing techniques have been intensively applied for the study of optical and microwave lenses, fiber optics, as well as reflectarray cells~\cite{ray_reflectarray1, Molero2021, ray1, ray2}. Moreover, ray tracing is extensively used for channel categorization in wireless communications, i.e.,~to estimate channel features of 5G networks~\cite{ray_propagationchannel1, ray_propagationchannel2} .

The \emph{transfer-matrix method} is a numerical tool for the computation of band structures (phase and attenuation constants) in periodic configurations. This method relays on the extraction of the so-called transfer matrix, which defines the properties of a single unit cell, and~the subsequent resolution of an eigenvalue problem~\cite{transfermatrix_mesa}. This method has proven to be valuable in many configurations, even in those involving complex metamaterials embedded in anisotropic media~\cite{transfermatrix_antonio}. In~order to extract the transfer matrix, different approaches can be used. Some works take advantage of general-purpose commercial software to extract the transfer (or scattering) matrix. This would be the case of~\cite{transfermatrix_example2}, where a 3-D configuration consisting in a corrugated dielectric waveguide with absorbing layers is analyzed. Simiarly, a~double wire-mesh structure that suppresses the cut-off frequency is analyzed in~\cite{transfermatrix_oscar}. Other works implement in-house techniques to extract the transfer matrix. This is the case exposed in~\cite{Transfer3D}, where a transfer-matrix method based on a plane-wave expansion is developed for the computation 3-D photonic platforms (i.e., a~woodpile structure).

Table~\ref{table_numerical} presents a summary of the main features related to all discussed computational techniques. The~strengths of each method are marked in bold~type.

\begin{table}[H]
        \caption{Strengths and weaknesses of the discussed computational techniques applied to 3-D design. Legend: ``L'' states for low, ``M'' states for medium, and ``H'' states for~high.}
	\setlength{\tabcolsep}{2.7mm}
	
		\label{table_numerical}
\begin{adjustwidth}{-\extralength}{0cm}
\centering 
\begin{tabular}{cC{2cm}C{2cm}ccc}\toprule
			\multirow{2}{*}{{\textbf{Methods}}} & 
			{\textbf{Generality,}} {\textbf{Robustness}}
			 &			{\textbf{Mathematical 
			Complexity}}& 
			\multirow{2}{*}{{\textbf{Computational 
			Complexity}}}&
			\multirow{2}{*}{{\textbf{Accuracy}}}&
			\multirow{2}{*}{{\textbf{Physical
			Insight}}}
\\ \midrule

			{Finite Differences} & 
			\begin{tabular}[c]{@{}c@{}} 
			{\textbf{M-H
}}
			\end{tabular} &
			\begin{tabular}[c]{@{}c@{}} 
			{M}
			\end{tabular} & 
            \begin{tabular}[c]{@{}c@{}} 
			{M-H}
			\end{tabular} &
			\begin{tabular}[c]{@{}c@{}} 
			{\textbf{M-H}}
			\end{tabular} &
			\begin{tabular}[c]{@{}c@{}} 
			{L}
			\end{tabular}
		
\\\midrule

			{Finite Elements} & 
			\begin{tabular}[c]{@{}c@{}} 
			{\textbf{H}}
			\end{tabular} &
			\begin{tabular}[c]{@{}c@{}} 
			{H}
			\end{tabular} & 
            \begin{tabular}[c]{@{}c@{}} 
			{H}
			\end{tabular} &
			\begin{tabular}[c]{@{}c@{}} 
			{\textbf{H}}
			\end{tabular} &
			\begin{tabular}[c]{@{}c@{}} 
			{L}
			\end{tabular}
		
\\\midrule

			{Integral Equations} & 
			\begin{tabular}[c]{@{}c@{}} 
			{\textbf{M-H}}
			\end{tabular} &
			\begin{tabular}[c]{@{}c@{}} 
			{M-H}
			\end{tabular} & 
            \begin{tabular}[c]{@{}c@{}} 
			{M-H}
			\end{tabular} &
			\begin{tabular}[c]{@{}c@{}} 
			{\textbf{M-H}}
			\end{tabular} &
			\begin{tabular}[c]{@{}c@{}} 
			{M}
			\end{tabular}
		
\\\midrule
	
			{Modal Analysis} &
			\begin{tabular}[c]{@{}c@{}} 
			{M}
			\end{tabular} &
			\begin{tabular}[c]{@{}c@{}} 
			{M-H}
			\end{tabular} & 
            \begin{tabular}[c]{@{}c@{}} 
			{M}
			\end{tabular} &
			\begin{tabular}[c]{@{}c@{}} 
			{\textbf{M-H}}
			\end{tabular} &
			\begin{tabular}[c]{@{}c@{}} 
			{\textbf{M-H}}
			\end{tabular}
		
\\\midrule

            {Circuits} &
			\begin{tabular}[c]{@{}c@{}} 
			{L-M}
			\end{tabular} &
			\begin{tabular}[c]{@{}c@{}} 
			{\textbf{L-M}}
			\end{tabular} & 
            \begin{tabular}[c]{@{}c@{}} 
			{\textbf{L}}
			\end{tabular} &
			\begin{tabular}[c]{@{}c@{}} 
			{L-M}
			\end{tabular} &
			\begin{tabular}[c]{@{}c@{}} 
			{\textbf{H}}
			\end{tabular}
		
\\\midrule

            {Ray Optics} &
			\begin{tabular}[c]{@{}c@{}} 
			{L-M}
			\end{tabular} &
			\begin{tabular}[c]{@{}c@{}} 
			{\textbf{L-M}}
			\end{tabular} & 
            \begin{tabular}[c]{@{}c@{}} 
			{\textbf{L-M}}
			\end{tabular} &
			\begin{tabular}[c]{@{}c@{}} 
			{L-M}
			\end{tabular} &
			\begin{tabular}[c]{@{}c@{}} 
			{\textbf{M-H}}
			\end{tabular}
		
\\\midrule

            {Homogenization} &
			\begin{tabular}[c]{@{}c@{}} 
			{L-M}
			\end{tabular} &
			\begin{tabular}[c]{@{}c@{}} 
			{L-H}
			\end{tabular} & 
            \begin{tabular}[c]{@{}c@{}} 
			{\textbf{L-M}}
			\end{tabular} &
			\begin{tabular}[c]{@{}c@{}} 
			{L-M}
			\end{tabular} &
			\begin{tabular}[c]{@{}c@{}} 
			{\textbf{M-H}}
			\end{tabular}
		
\\\midrule

            {Transfer Matrix} &
			\begin{tabular}[c]{@{}c@{}} 
			{M}
			\end{tabular} &
			\begin{tabular}[c]{@{}c@{}} 
			{\textbf{L-M}}
			\end{tabular} & 
            \begin{tabular}[c]{@{}c@{}} 
			{M}
			\end{tabular} &
			\begin{tabular}[c]{@{}c@{}} 
			{M}
			\end{tabular} &
			\begin{tabular}[c]{@{}c@{}} 
			{M}
			\end{tabular}
		
\\\bottomrule

	\end{tabular}	
\end{adjustwidth}
\end{table}


\subsection{Commercial~Solvers}
Commercial simulators are robust general-purpose tools employed for the resolution of a multitude of engineering and physics problems. Their development has been enormous, and these days it is normal to find them embedded in a complete suite that includes many types of solvers that adapt to the simulation requirements: frequency/time domain, large/finer mesh, etc. These solvers are normally based in the methods previously discussed in this~section.

In the field of electromagnetics, three of the most popular commercial simulators are CST Microwave Studio~\cite{CST}, Ansys HFSS~\cite{hfss}, and COMSOL Multiphysics~\cite{comsol}. CST and HFSS share some similarities in their internal design. Both have a complete suite of solvers. In~CST, the~frequency-domain solver employs a FEM method, the time-domain solver  employs a finite integration technique (a generalization of the FDTD method), the integral-equation solver combines a standard MoM with a fast multipole method (FMM), and a ray-tracing solver is available for very large structures. Similarly, the~frequency-domain, integral-equation, and~ray-tracing solvers in HFSS employ FEM, MoM, and ray-theory methods, while the time-domain solver is based in a FEM in this case. On~the other hand, COMSOL Multiphysics is primarily based on FEM formulations that make use of either direct (LU decomposition) or iterative~solvers.

\section{Designs}\label{sec3}

  A large number of functionalities can be attributed to 3-D metamaterials, as~can corroborated in the literature (mainly papers from the current century). It can be established that the beginning of metamaterials was linked to the experimental verification of negative permittivity and permeability in artificial materials~\cite{Pendry1999}. The~primary prototype of a metamaterial consisted of a 3-D periodic distribution of split-ring resonators, performing electrical properties not accessible in nature and~\emph{easily} tunable. Many researchers paid attention to these exotic properties, giving rise to the development of potential applications, such as~those in the field of magnetic resonances~\cite{Freire2008, Algarin2010}. The~3-D architecture of split-ring prototypes is otherwise difficult to model. It is worth remarking the homogenization technique reported in~\cite{Baena2008}, reducing the analysis of such a complicated architecture to simple equivalent circuits. As~mentioned in the previous section, homogenization techniques are well known and have encouraged the study of some other types of complicated 3-D structures~\cite{Silveirinha2009, mario3D}. Such is the case of artificial photonics crystals, reported by Dr. Sievenpiper~et~al. before the metamaterials \emph{boom} \cite{Sievenpiper1996}. {Artificial photonic crystals were motivated by the well-known propagating properties of photonic crystals, emulated, for~example, in~metasurfaces operating in the THz regime~\cite{Fietz2011}}, specifically wire mesh photonics crystals formed by fully metallic lattices of parallel rods~\cite{Belov2003} or interconnected wires under exotic geometries~\cite{comsol3D_2, comsol3D_1, Liang2016}. The~interest in these topologies especially lies in the inherent propagation properties, leading to anysotropy and effective-permittivity control~\cite{https:Simovski2012, Kushiyama2012}. Applications straightforwardly profiting these properties involve near-zero permittivity materials, directivity enhancement, or subwavelength~imaging. 
  

Successive evolutions and extensions of wire metamaterials have emerged and have been reported in the last decade. Three-dimensional designs based on cells including non-connected helical wires are noteworthy in exploiting chirality, {such as the one in~\cite{Gansel2009} or~that in~\cite{Faniayeu2020}. The latter considered helical elements with exotic orientations and combinations in order to enhance the chiral properties}. Chirality is a powerful tool providing polarization manipulation in the form of linear-to-circular conversion~\cite{Gansel2009, Wu2019_helice} or linear-to-linear field rotation~\cite{Wu2019_helice, Wu2019}. The~physical mechanism behind the field conversion profits from the inherent anysotropy of chiral elements, which manipulates the co- and cross-polar field components when the whole structure is fed by an impinging plane wave. {A quite interesting study of chirality regarding a set of cell candidates was reported in~\cite{Corbaton2019}}.  Additional prototypes based on periodic distributions of connected and non-connected wires have also been studied and reported for different purposes. Such is the case in~\cite{Ehrenberg2012}, where a narrow-band three-dimensional lens was constructed, or~the cases in~\cite{Sanz2014, Zhu2018} reporting periodic distributions of folded-wires with the aim of enhancing angular stability. It is finally worth remarking the very original designs conceived by the group of Prof. Mittra in the last decade, turning the screw of the geometry in wire metamaterials~\cite{Mittra2013, Pelletti2013, Pelletti2013_2}. The~unit cell comprises two metallic Jerusalem crosses separated by a certain distance and connected via bifilar lines. This configuration is fed by external plane waves, exhibiting wideband response with angular~robustness. 

An alternative and very important family of 3-D periodic structures has mainly been developed by Prof. Shen and his research group. Their designs can easily be recognized by a particular physical architecture, and their optimization techniques are based on well-constructed multi-modal mathematical approaches. A~first review was published in 2014 in~\cite{Rashid2014}. The~interest in these architectures lies in the large number of functionalities that can be invoked thanks to, among~other reasons, the~introduction of resonators extended along the propagation direction. The~orientation of the resonators breaks up the axial symmetry typical from 2-D periodic arrays, adding new degrees of freedom (geometrical parameters) to the~structures.

As can be checked in their papers, the~general prototype proposed by them are two-dimensional periodic arrays formed by unit cells emulating a sort of microstrip geometry. See, for~example,~\cite{Omar2016} or~\cite{Omar2017tap}, where cells with similar appearance are used for different functionalities. In~particular, the~cells in~\cite{Omar2016, Wang2019_bandstop} are employed to exhibit bandstop operation. This physical mechanism behind bandstop response is mainly governed by strip and dog-bone-shaped resonators included along the microstrip and is~better understandable thanks to the derivation of equivalent circuits. The~use of circuits was also necessary to invoke the inverse operation (bandpass) in the original cell reported in~\cite{Omar2019_bandpass}, intended to be used for radome applications. It is worth remarking the complexity encountered to optimize 3-D designs in transmission operation. The~cell presented in~\cite{Li2020} is an example, where stepped-slits along the axial direction govern the transmission properties of the whole device, but~only single-pol operation is possible. A~similar configuration having dual-pol operation is presented in~\cite{Li2019}, providing a circular polarizer for the~Ku-band.       

\begin{figure}[H]
    \includegraphics[width= 0.9\textwidth]{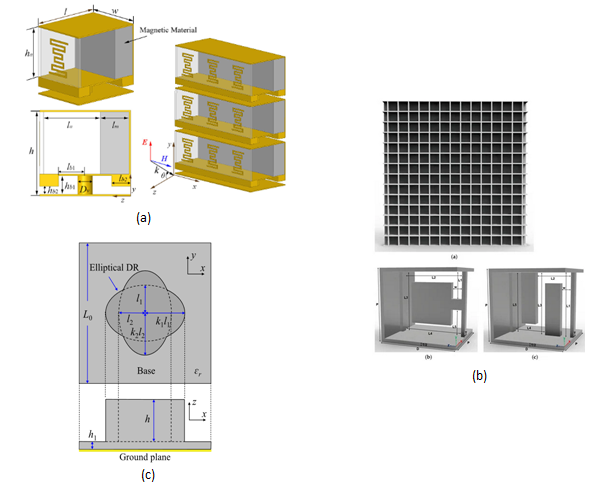} 
    \caption{(\textbf{a}): 
 One of the cells conceived for rasorbers. This one was reported in
~\cite{Wang2021}. (\textbf{b}): Full-metal reflectarray prototype reported in~\cite{Velasco2021}. (\textbf{c}): Fully dielectric reflectarray prototype reported in~\cite{Sun2021}. } 
	\label{CeldasDesigns}
\end{figure}

The \emph{pseudo}-microstrip geometry has otherwise been highly exploited for applications involving radar cross-section reduction, low observability, and multi-channel selection. This is possible thanks to the efficient combination between the inherent wideband and multiband nature of the cell with resistive elements. The~most straightforward design was presented in~\cite{Omar2017tap} after combining the cell-type in~\cite{Omar2016} with commercial resistors. The~result was a dual-pol wideband absorber experimentally tested with success. Other variants of wideband absorbers were reported in~\cite{Luo2020, Omar2017}, where the unit cell comprises long folded strips. Multi-path cells constitute a third variant of absorbers, in which the operation band is controlled by resistive pin-diodes~\cite{Zhou2021}. 
Ultrawideband absorbers can also be used to design absorptive frequency-selective transmission (AFST) \cite{Omar2018, Omar2020} or frequency-selective reflection structures (AFSR) \cite{Yu2019, Yu2019_hybrid}. The latter are better known as \emph{rasorbers}. The~main property of both models is the characteristic spectral response, which introduces a transmission/reflection band in the middle of a wide absorption band. In~other words, the~devices are configured to transmit/reflect incoming waves of which the frequency is inside a particular band and~discriminate (absorb) incoming waves with the rest of frequencies. Unit cells with multi-path configurations based on exotic of designs of microstrips are again employed. Specifically, the~strategy consists of designing cells having artificial and parallel paths (transmission lines connected in series), where some of them are \emph{resistive} pathways for the traveling waves and~the rest stay lossless~\cite{Yu2017, Tianwei2017}. This combination, when properly optimized, introduces sequences or concatenations of transmission/reflection and absorption bands. The~resistive pathways can be constructed by the use of magnetic materials~\cite{Tianwei2017, Wang2021, Huang2020}. They are materials artificially designed to have larger permeability loss tangent. The~propagation along them is highly lossy, contributing to an efficient absorption. A~visual example of a cell of this kind is sketched in Figure~\ref{CeldasDesigns}a. Other ways to design resistive pathways are the introduction of commercial ferrites~\cite{Wang2020} or commercial resistors~\cite{Yu2017, Li2014}. 

Periodic distributions of waveguide-cell configurations constitute another great family of 3-D structures. Waveguide-cell topologies involve fully metallic, fully dielectric, or metallo-dielectric designs with square cross section. Naturally, they are generally complemented with the introduction of resonators along the axial direction with the aim of profiting from a bigger number of degrees of freedom. It is noteworthy to begin with a first sub-family labeled as \emph{square-coaxial} waveguides~\cite{Ge2016, Tang2016, Zhu2018}. The~cell is formed by concentric dielectric and metallic square rings, all having the same length. The~dielectric materials are commonly commercial resins and plastics suitable for 3-D-printing fabrication. This type of cell encourages multi-channel transmission, with~each single channels being that enclosed in the region between two adjacent metallic rings. As~was aforementioned, multi-paths encourage multi-band response (multi-band filters). In~this case, the~paths are well isolated from each other, giving rise to highly isolated bands with no mutual interference. Furthermore, strong robustness with the incidence angle was achieved. 
A second version of coaxial waveguide-cells includes free-space channels. This is the case reported in~\cite{Zhu2018coaxial, Zhu2019}. The~free-space channel is actually a square hole in the center of the unit cell. It enhances the generation of transmission poles and zeros, encouraging dual-band transmission that can be easily tuned. 

Full-metal 3-D periodic structures are part of a novel and modern 3-D family. Its development was recent due to the historically high difficulty in achieving mechanically robust and self-supported devices manufactured by metal only. The~fast increase in 3-D-printing techniques in recent decade has allowed us to address the fabrication of this type of architectures, though~its use is not widespread yet. A~massive use is expected in the near future due to future trends and new considerations introduced in recent communications standards, highly keen to cover millimeter bands. Spacial applications are also demanding fully metallic terminals owing to the hard conditions in spacial environments. Dielectric materials are quite limited in these scenarios, from~the electromagnetic, mechanical, and thermal point of view. Additionally, dielectric losses become dominant in the far-microwave spectral region and beyond, drastically deteriorating the performance of the devices under use. All of these constraints motivate the search, design, and optimization of 3-D fully-metallic architectures. First attempts have been published in~\cite{Tang2016} and were later refined in~\cite{molero_circuit3D, Molero2020} with relative success. Fully metallic cells are the basis of the periodic structure, consisting of square waveguides with resonators perforated on the walls. Thanks to the resonators, the~control of the transmission features, such as~the transmission-band positions and dual-polarization operation, is readily tuned. These properties were successfully synthesized to attain single and dual-pol circular polarizers in~\cite{molero_circuit3D, Molero2020}, respectively. The~same principles were later applied for the design, optimization, and fabrication of the reflectarray~\cite{Velasco2021} shown in Figure~\ref{CeldasDesigns}b. Furthermore, this same structure is currently the subject of study for a most challenging terminal, which would include reconfiguration with graphene, leading to a metastructure exhibiting several functionalities~\cite{Molero2021}. {The latest version of this structure was recently published~\cite{molero3D_2022} and reported slits instead of resonators on the walls. This new architecture is suitable for wideband transmission, since the TEM mode can be excited in the cell. A~wideband circular polarizer has theoretically been predicted and experimentally tested with success.}         

Finally, we highlight several 3-D prototypes specifically designed and manufactured for a fixed functionality. Such are the cases of 3-D reflectarrays/transmitarrays and 3-D lenses. In~both cases, the~use of periodic structures is no longer possible due to the need to implement phase gradients all along the surfaces. Since the cell distribution is considered \emph{pseudo} periodic, we use the term \emph{metasurface}. {In the last decade, the~use of metasurfaces has increased worldwide, thanks to the numerous theoretical predictions that have been corroborated experimentally, and in~application fields where they have become essential~\cite{Bukhari2019}}. Metasurfaces are low-profile terminals constituted by cells in which the spacial dimensions do not overcome the operation wavelength (in contrast with multi-layer versions of 2-D periodic metasurfaces). This makes them highly attractive, especially when terminals with small sizes are required. For~instance and~as mentioned above, spacial applications demand low-weight, low-profile and fully-metallic devices for future missions. Fully metallic reflectarrays are still challenging, but~some prototypes have been launched. This is the case in~\cite{Velasco2021} (previously cited) and in~\cite{Mei2020}. This latter is quite interesting because the reflectarray was mechanically reconfigured and optimized to operate in the Ka-band. Furthermore, it has a privileged architecture to provide wideband operation in frequency. A~fully metallic transmitarray was presented in~\cite{Wang_RA_2021} for the Ka-band, with~wideband response and excellent insertion losses. Examples regarding fully dielectric reflectarrays do also exists in the literature. It is worth mentioning that the term fully dielectric is generally not accurate owing to the inclusion of ground planes that forces the fully reflection operation. In~this sense, dielectric reflectarrays based on dielectric substrate with protrusions are the most appealing designs. Paradigmatic examples are reported in~\cite{Sun2021} with elliptical protrusions (see Figure~\ref{CeldasDesigns}c for visual information) \cite{Li2020_RA} with cross-shaped protrusions and~\cite{Yang2014} with rectangular protrusions. The~protrusions are oriented in order to break the symmetry of the cell, with~the role of controlling the vertical and horizontal field components in a context of high but not full independent. This kind of reflectarray provide very wideband response in frequency. Another version of a dielectric reflectarray was published in~\cite{Mei2020bis}, where the protrusion (C-shaped in this case) was not supported on a dielectric panel but on the ground plane directly. The~properties are actually similar to the dielectric reflectarrays previously cited. A~final version of a reflectarray is that called \emph{Kirigami} \cite{Cui2020}, where an effective conformal shape just varies the size of cells. In~this particular structures, the cells have a sort of hexagonal shape. In~the context of 3-D lenses, it is worthy to highlight the prototypes reported in~\cite{PrintedLens1, PrintedLens2}, consisting of \emph{Luneburg} and \emph{Fresnel} lenses, respectively. The~first of them comprises exotic cubic cells to design the lens in the Ku/K-bands, whereas the second one combines air and dielectric slabs as a way to control the phase of each cell for optimization of the lens in the THz domain. Both cases shows high efficiency and very good gain~performances.

\section{Fabrication~Techniques}\label{sec4}

Due to the 3-D geometry of the unit cells that compose the metamaterial under studied in this review, new manufacturing techniques and assemblies have had to be employed in order to achieve a satisfactory prototyping. In~planar conventional fabrication where metasurfaces offered 2-D geometry, the~PCB manufacturing achieved adequate prototyping even when metasurfaces were used to form stacks. However, for~the case of unit cell fabrication with 3-D geometry, these results are inadequate and, if used, have to be complemented by more complex assembly. With~the development of new manufacturing techniques in \emph{3-D printing}, the~task of manufacturing metamaterials with 3-D geometry has been alleviated, allowing for their manufacture with less complexity and higher precision. This also means that, with 3-D printing, prototypes can be produced in a single piece without having to divide the structure into different pieces, to manufacture each one separately, and then to assemble them, which results in a more tedious process and a greater potential source of errors. In~this section, a~revision of different works in the literature is carried out, which show a prototype of a unit cell with 3-D geometry. Generally speaking, these can be divided into three types of manufacturing: (1) conventional with 3-D assembly, (2) 3-D printing, and (3) alternative~techniques.

\subsection{Conventional Manufacturing Techniques and 3-D~Assembly}

The first one to be described is conventional manufacturing  with \emph{3-D assembly}. As~mentioned above, this fabrication technique can be represented by any traditional manufacturing technique such as PCB manufacturing, computer numerical control (CNC), or electrical discharge machining (EDM), which entails a separate fabrication of the parts that conform the prototype and, later, an~assembly of these parts in three dimensions. Examples of this kind of prototyping are found in~\cite{Omar2016,Li2019,Zhu2019,Wang2019_bandstop}. In~all of these studies, the~3-D unit cells have been manufactured with different PCB layers stacked both horizontally and vertically, forming the sides of the unit cells. To~ assemble these pieces in a simple way, different grooves are implemented to interlock them all together (see Figure~21c from~\cite{Wang2019_bandstop}). In~addition to this mechanical assembly, most of these designs require the application of conductive glue~\cite{Zhu2019} or solder~\cite{Wang2019_bandstop,Li2019} in the assembly areas (grooves) to avoid any electrical discontinuity that could damage the electromagnetic behavior of the manufactured prototype. An~alternative for 3-D assembly with PCB layers is the one presented in~\cite{Pelletti2013}, where the connecting and supporting elements between the two layers are copper wires, which in turn have a beneficial effect on the electromagnetic response. Thanks to the 3-D geometry of the unit cells, lumped elements or homogeneous materials with a given dielectric properties are relatively easy to insert into the unit cells. Some examples in 3-D PCB assembly are shown in~\cite{Omar2017tap,Luo2020,Wang2021} and Figure~\ref{CeldasDesigns}a illustrates a unit cell where magnetic material is included in its structure. In~general, due to the current applications, these external elements are lossy such as resistors or magnetic/dielectric materials to design absorbers~\cite{Luo2020} or rasorbers~\cite{Omar2017tap,Wang2021}. However, in~these last two studies, additional metal structures (for example, conducting planes) had to be fabricated separately to allow for the integration of the lumped elements or the lossy materials. On~the other hand, a~conventional manufacturing technique EDM has also been employed to achieve an all-metal metamaterial. One of the example that utilizes this manufacturing technique is the prototype presented in~\cite{Molero2020}. Thanks to the periodic slot geometry along the horizontal and vertical walls of the waveguides that form the unit cells, in~this case, there is no need for 3-D assembly and the prototype is in a single piece. As~mentioned in this work, CNC has been discarded as it required a more complex assembly~process.

\subsection{3-D~Printing}

\begin{figure}[H]
\hspace*{2cm}
    \subfigure[]{\includegraphics[width= 0.25\textwidth]{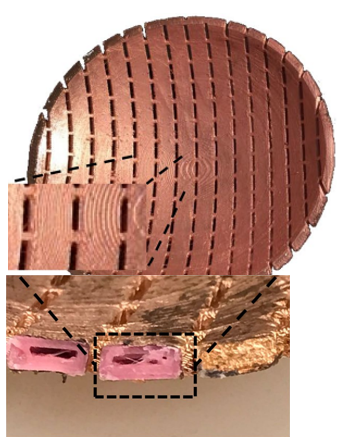}
	}
    \subfigure[]{\includegraphics[width= 0.49\textwidth]{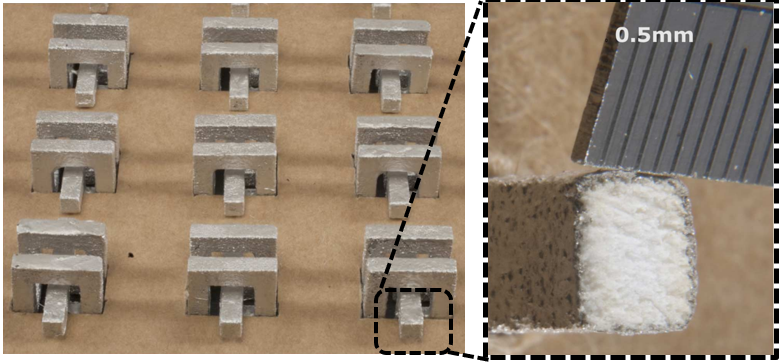}
	}
    \caption{Examples of 3-D metamaterials manufactured by 3-D printing with subsequent metallization. (\textbf{a}) Design from
~\cite{Fernandez_Alvarez2021} fabricated using FDM and electroplating. (\textbf{b}) Design from~\cite{Sanz2014} fabricated using MJ and silver~coating.} 
	\label{Fig:Metalizacion}
\end{figure}

The second type of manufacturing described in this section is one that involves 3-D printers. Within~this category, a~distinction can be made between prototypes that are 3-D printed on \emph{plastic} with subsequent metallization and~prototypes that are 3-D printed \emph{directly on metal}, which generally do not require subsequent metallization. First, we revisit the works that employ a 3-D printing in plastic and then apply a conductivity layer all over their surface. This manufacturing option is the most common in the literature for the fabrication of 3-D metamaterials. Several examples of this type of 3-D manufacturing can be seen in the prototypes of the following~\cite{Ehrenberg2012,Sanz2014,Tang2016,Liang2016,Ishikawa2017,Zhu2018,Sadeqi2019,Kien2020,comsol3D_1,Velasco2021,Fernandez_Alvarez2021,Zhang2021,Vasquez2022}. Different technologies related to plastic 3-D printing are also available and some of them were used in previous work. Fused filament fabrication or \emph{fused deposition modeling} (FDM) is the most common for general-purpose 3-D printers. It is based on the fusion of a thermoplastic in which the extruder nozzle produces the printed piece. Prototypes presented in~\cite{Ishikawa2017,Fernandez_Alvarez2021} use 3-D printers based on FDM in which the thermoplastic filaments are \emph{polylactic acid} (PLA) and \emph{acrylonitrile butadiene styrene} (ABS). Figure~\ref{Fig:Metalizacion}a shows the prototype in the work by~\cite{Fernandez_Alvarez2021}, where a cross-section of the unit cell clearly shows how it was manufactured. Another 3-D printing technology presented in the above studies is the one based on \emph{material jetting} (MJ), in which the principle of operation is based on the deposition of material droplets in each layer that are cured by UV light before another layer is printed. This type of 3-D printing technique is used in~\cite{Ehrenberg2012,Sanz2014,Zhu2018}, and Figure~\ref{Fig:Metalizacion}b reveals the cross-section of the unit cell reported in~\cite{Sanz2014}. Similar to the previous 3-D printing technology, it is the 3-D printing based on \emph{stereolithography} (SLA). The~similarity lies in the use of UV light to cure the material (in this case, resin) layer by layer. However, the~difference between both of them is that the SLA makes the cure in a tank full of resin while MJ cures the material after the material has been sprayed. Unit cell designs that have employed SLA in their prototyping are presented in~\cite{Tang2016,Liang2016,Sadeqi2019,Velasco2021,Zhang2021,Vasquez2022}. The~last technology applied for the manufacture of 3-D metamaterials using 3-D printing on plastic is called \emph{selective laser sintering} (SLS) used in~\cite{Kien2020,comsol3D_1}. In~contrast with the previous 3-D printing technologies, this one sinters and fuses a preheat polymer powder by a laser to create the printed piece. The latter uses nylon as the sinterized material in the 3-D printing process. Since the 3-D printing technologies described above print the unit cells on plastic and they have conceived to be metallic, a~metallization process is needed to coat the surface of the printed unit cells with a highly conductive metal. Different techniques to metallize the surface of the unit cells have been seen in previous work: brush paint~\cite{Zhu2018,Kien2020}, spray~\cite{Velasco2021}, electroplating~\cite{Fernandez_Alvarez2021,Vasquez2022}, electroless plating~\cite{Ishikawa2017}, sputtering~\cite{Ehrenberg2012,Sadeqi2019}, dip coating~\cite{comsol3D_1}, and~vacuum chamber metallization~\cite{Zhang2021}. Metallization is a crucial part of this manufacturing process as each of the above techniques offers a level of roughness, conductivity, and layer thickness that must be taken into account depending on the operating frequency range and geometry of the unit cell. It is important to mention that 3-D printing on plastic has been also extensively employed in fully dielectric unit cells in the lens design~\cite{PrintedLens1,PrintedLens2}. To~end with 3-D printing techniques, there exists an alternative 3-D printing process that prints directly on metal and, thus, avoids the metallization process. {Few studies in the literature have employed this 3-D printing technique. In~\cite{Wu2019_helice}, a~3-D metamaterial printed on a cobalt–chromium alloy by \emph{selective laser melting} (SLM) was presented. In~contrast to the previous work, in~\cite{molero3D_2022}, the~same 3-D printing technique was used but the printed metal was titanium. The~choice of this type of metal instead of other metals such as aluminum alloys lies in the fact that it is possible to achieve a higher precision in the metal printing. Nevertheless, SLM needs high control and power in the laser since it has to reach the melting point of the printed metal, contrary to the similar 3-D printing technique named \emph{direct metal laser sintering} (DMLS), where the metal powder is preheated and the laser power does not need to be as high as in SLM.}

\unskip

\subsection{Alternative Techniques for 3-D~Prototyping}

Although 3-D assembly and 3-D printing techniques have proven to be very successful, other manufacturing alternatives can be also found in the literature to produce 3-D metamaterials. In~\cite{Harnois2020}, \emph{water transfer printing} (WPT) technology was employed to successfully transfer a metasurface into a 3-D metal mold. Before~using the WPT, the~metasurface was manufactured by \emph{inkjet printing} on a flexible substrate such as polyethylene terephthalate (PET). With~the use of WPT, the~unit cells achieve a 3-D shape due to the mold geometry. The~design presented in~\cite{Nauroze2018} produced a 3-D FSS constituted by folded dipoles that are inkjet printed on a cellulose paper. This was selected as a flexible substrate of the prototype because it can produce origami patterns that are inherently 3-D structures. {Instead of using flexible substrates, as~in the previous work, in~\cite{Cho2018}, FDM was used to create a 3-D structure, and then, cross dipoles and square loops were printed with metallic ink.} For 3-D metamaterial that were intended to operate at terahertz frequencies, nanometer-precision manufacturing techniques were needed for functional prototyping. The work in~\cite{Burckel2010,Xiang2013,Pan2021} used fabrication techniques based on \emph{lithography} to achieve these levels of manufacturing precision. As~a summary, Table~\ref{tab:TableManufact} presents all of the papers presented in this~section.

\begin{table}[H]
		\caption{{Summary of the fabrication characteristics of the studies mentioned in Section~\ref{sec4}.}}
		
		\label{tab:TableManufact}
		\setlength{\tabcolsep}{2.4mm}

\begin{adjustwidth}{-\extralength}{0cm}
\centering 
\begin{tabular}{cccc}\midrule
			{\textbf{Works}}          
			& \begin{tabular}[c]{@{}c@{}} 
			{\textbf{Fabrication Technique}}
			\end{tabular} & 
			\begin{tabular}[c]{@{}c@{}} 
			{\textbf{3-D Assembly}}
			\end{tabular} & 
            \begin{tabular}[c]{@{}c@{}} 
			{\textbf{Required Metallization}}
			\end{tabular}

			\\ \midrule
			
			{\begin{tabular}[c]{@{}c@{}} 
			{\cite{Omar2016,Li2019, Zhu2019, Wang2019_bandstop, Pelletti2013, Omar2017tap, Luo2020, Wang2021}
 }
			\end{tabular}} 
			 & \begin{tabular}[c]{@{}c@{}} 
			{PCB fabrication}
			\end{tabular} & 
			\begin{tabular}[c]{@{}c@{}} 
			{Yes}
			\end{tabular} & 
            \begin{tabular}[c]{@{}c@{}} 
			{No {*}}
			\end{tabular}
			
 \\ \midrule
 
			{\begin{tabular}[c]{@{}c@{}} 
			{ \cite{Molero2020} }
			\end{tabular}} 
			 & \begin{tabular}[c]{@{}c@{}} 
			{EDM}
			\end{tabular} & 
			\begin{tabular}[c]{@{}c@{}} 
			{No}
			\end{tabular} & 
            \begin{tabular}[c]{@{}c@{}} 
			{No}
			\end{tabular}
			
 \\ \midrule
 
			{\begin{tabular}[c]{@{}c@{}} 
			{ \cite{Ishikawa2017,Fernandez_Alvarez2021} }
			\end{tabular}} 
			 & \begin{tabular}[c]{@{}c@{}} 
			{FDM 3-D printing}
			\end{tabular} & 
			\begin{tabular}[c]{@{}c@{}} 
			{No}
			\end{tabular} & 
            \begin{tabular}[c]{@{}c@{}} 
			{Yes}
			\end{tabular}
			
 \\ \midrule
 
			{\begin{tabular}[c]{@{}c@{}} 
			{ \cite{Ehrenberg2012,Sanz2014,Zhu2018} }
			\end{tabular}} 
			 & \begin{tabular}[c]{@{}c@{}} 
			{MJ 3-D printing}
			\end{tabular} & 
			\begin{tabular}[c]{@{}c@{}} 
			{No}
			\end{tabular} & 
            \begin{tabular}[c]{@{}c@{}} 
			{Yes}
			\end{tabular}
			
\\ \midrule
			
			{\begin{tabular}[c]{@{}c@{}} 
			{ \cite{Tang2016,Liang2016,Sadeqi2019, Velasco2021,Zhang2021,Vasquez2022} }
			\end{tabular}}
			 & \begin{tabular}[c]{@{}c@{}} 
			{SLA 3-D printing}
			\end{tabular} & 
			\begin{tabular}[c]{@{}c@{}} 
			{No}
			\end{tabular} & 
            \begin{tabular}[c]{@{}c@{}} 
			{Yes}
			\end{tabular}
			
\\ \midrule

            {\begin{tabular}[c]{@{}c@{}} 
			{ \cite{Kien2020,comsol3D_1} }
			\end{tabular}} 
			 & \begin{tabular}[c]{@{}c@{}} 
			{SLS 3-D printing}
			\end{tabular} & 
			\begin{tabular}[c]{@{}c@{}} 
			{No}
			\end{tabular} & 
            \begin{tabular}[c]{@{}c@{}} 
			{Yes}
			\end{tabular}
			
\\ \midrule

            {\begin{tabular}[c]{@{}c@{}} 
			{ \cite{Wu2019_helice,molero3D_2022} }
			\end{tabular}} 
			 & \begin{tabular}[c]{@{}c@{}} 
			{SLM 3-D printing}
			\end{tabular} & 
			\begin{tabular}[c]{@{}c@{}} 
			{No}
			\end{tabular} & 
            \begin{tabular}[c]{@{}c@{}} 
			{No}
			\end{tabular}
			
\\ \midrule

	        {\begin{tabular}[c]{@{}c@{}} 
			{ \cite{Harnois2020} }
			\end{tabular}} 
			 & \begin{tabular}[c]{@{}c@{}} 
			{Inkjet printing and WPT}
			\end{tabular} & 
			\begin{tabular}[c]{@{}c@{}} 
			{No}
			\end{tabular} & 
            \begin{tabular}[c]{@{}c@{}} 
			{No}
			\end{tabular}
			
\\ \midrule
			{\begin{tabular}[c]{@{}c@{}} 
			{ \cite{Nauroze2018} }
			\end{tabular}} 
			 & \begin{tabular}[c]{@{}c@{}} 
			{Inkjet printing on  origami structures}
			\end{tabular} & 
			\begin{tabular}[c]{@{}c@{}} 
			{No}
			\end{tabular} & 
            \begin{tabular}[c]{@{}c@{}} 
			{No}
			\end{tabular}
			
\\ \midrule
			{\begin{tabular}[c]{@{}c@{}} 
			{ \cite{Cho2018} }
			\end{tabular}} 
			 & \begin{tabular}[c]{@{}c@{}} 
			{Inkjet printing and  FDM}
			\end{tabular} & 
			\begin{tabular}[c]{@{}c@{}} 
			{No}
			\end{tabular} & 
            \begin{tabular}[c]{@{}c@{}} 
			{No}
			\end{tabular}
			
\\ \midrule

		    {\begin{tabular}[c]{@{}c@{}} 
			{ \cite{Burckel2010,Xiang2013,Pan2021} }
			\end{tabular}} 
			 & \begin{tabular}[c]{@{}c@{}} 
			{Lithography}
			\end{tabular} & 
			\begin{tabular}[c]{@{}c@{}} 
			{No}
			\end{tabular} & 
            \begin{tabular}[c]{@{}c@{}} 
			{Yes}
			\end{tabular}
			
\\ \midrule

		\end{tabular}
\end{adjustwidth}
{\footnotesize  {*} It may require soldering or conductive bonding to ensure electrical continuity.}

	\end{table}


\section{Future~Trends}\label{sec5}

Thus far, we  presented an overview of the existing computational methods, designs, and manufacturing techniques related to 3-D metamaterials, particulary in for microwave and photonic regimes. Nonetheless, times are evolving and some of the formerly discussed ideas may become obsolete in the future while others may have incredible potential, {such as the application of artificial intelligence algorithms}. Here, we present our vision on future trends and challenges in the field of 3-D~metamaterials.

Computational complexity increases notably when considering 3-D metamaterials compared with traditional 1-D and 2-D metasurfaces. Furthermore, 3-D geometries are substantially more difficult to model than planar configurations.  As~a direct consequence, gaining physical insight into the electromagnetic behavior of the structure is becoming increasingly complex. Certainly, FEM and finite difference methods are very appreciated generalist approaches, as~they can deal with all types of geometries and materials, normally at the expense of higher computational resources and a lack of physical insight. Thus, those methods (or hybrids) that reduce computation times and, at~the same time, provide physical insight will attract much attention for the analysis of 3-D structures. This is the case of reduced-order models such as equivalent circuits, ray optics, and homogenization methods. Ray optics and homogenization techniques are not new to 3-D metamaterials, even though their capabilities have not been fully exploited. However, circuit models applied to 3-D structures are rarely found in the literature. It is a field to be developed due to its enormous potential. On~the other hand, modal techniques are intermediate solutions between purely numerical approaches and reduced-order models. They offer a good performance in most of the features highlighted in Table~\ref{table_numerical}. Actually, Floquet modal expansions (together with FDTD and homogenization methods) are the first techniques to be implemented for the analysis of \emph{spacetime metamaterials} (periodic configurations modulated in space and time)~\cite{spacetime1, spacetime2, spacetime3}, which is one of the current {hot topics in electromagnetism~\cite{spacetime4, spacetime5, spacetime_microwave}}. In~fact, current commercial software fail to consider time modulations in the computation, so in-house codes are needed to model spacetime structures. {Additionally, \emph{artificial intelligence} (AI) has supposed a breakthrough in electromagnetics and electronics in solving challenging problems (scattering, radar, channel parameter estimation, microwave imaging, remote sensing, etc.) involving a large number of variables and design constraints~\cite{ai1, ai2, ai3, ai4, ai5}. In~the near future, it is expected that AI techniques such as machine learning or deep neural networks coexist with the more traditional computational methods reviewed in this document. }

In the context of cell designs and potential functionalities in the short and intermediate term, a~clear trend is determined by 5G and 6G communication standards. Specially the leverage of millimeter bands, where terminals must adequately be adapted to the operation wavelength. The~use of dielectric materials becomes problematic at these frequencies due to the strong presence of losses. The~use of 3-D fully metallic devices is expected to be widespread, as~long as the contemporaneous fabrication techniques correspondingly evolved. This need for full-metal structures is shared by spatial-applications demands~\cite{Chahat2018}, since spacial environments are exposed to very extreme conditions (thermal, electromagnetics, etc.). At~the same time, 3-D metamaterials allows for a \emph{perfect} decoupling between vertical and horizontal polarizations. This motivates the search for and improvement in functionalities and applications where the control of both field components must be carefully controlled. Such is the case of circular polarizers, polarization converters, absorbers, etc. The design of dual-pol absorbers, in~particular those called rasorbers for radar scenarios, deserve special attention. The~independent control of both polarizations would induce reflection/absorption independent channels for each of them.  
Three-dimensional structures are also incorporated into the field of \emph{reconfigurable intelligent surfaces} (RIS), where the strict control of field patterns emitted by the metasurface is realized via reconfiguration. Despite the bulk of research relying on a reconfigurable material (especially to operate at millimeter bands without loss of efficiency), the~effective number of field patterns increases when both polarizations are relevant instead of one. Three-dimensional metasurfaces are therefore good candidates for future RIS~\cite{Molero2021} and the promising contexts where RIS will be applied, e.g., the intelligent vehicle~\cite{matthaiou2020} and vortex-generation for RF imaging systems~\cite{Chen2020}, etc.    

Regarding the manufacturing techniques employed for the production of 3-D metamaterials, Table~\ref{tab:TableManufact} clearly shows that the two that are mainly used are PCB fabrication with a 3-D assembly and 3-D printing in plastic (FDM, MJ, SLA, and SLS) with subsequent metallization. PCB fabrication has been one of the most developed manufacturing techniques in recent decades due to its wide use in integrated circuits. Nevertheless, the~designs prototyped with PCB technique always require a 3-D assembly and the presence of a substrate, which significantly increase the dielectric losses at millimeter-wave frequencies. On~the other hand, due to the great recent development of 3-D printing in recent years, a~great diversity of high-precision and cost-effective 3-D printers can be found. As~a consequence, 3-D printing is intensively applied nowadays for the production of high-frequency RF prototypes. However, it has the disadvantage of requiring metallization, which becomes complex for small details and low roughness. This is evidenced by the diversity of metal plating techniques used in the studies mentioned in Section~\ref{sec4} related to 3-D printing on plastics. Therefore, it is necessary to further improve the abovementioned manufacturing techniques. One of the envisaged solutions is the possibility of low-cost 3-D printing in metals, producing a low level of roughness and a good conductivity desired for 3-D metamaterials at millimeter-wave regime. These characteristics in the manufacturing would eliminate the need for metallization, even though a subsequent metallization is sometimes required to either increase the conductivity of the base metal or to minimize the surface roughness of the printed prototype~\cite{RoughnessSLM}. 

\section{Conclusions}\label{sec6}
In this paper, we presented an overview of the past, present, and future challenges related to 3-D metamaterials. We introduced the main limitations that have prevented 3-D metamaterials to be properly utilized in the past, as~well as the advantages that 3-D configurations can offer compared with their 1-D and 2-D~counterparts.

First, we reviewed the most relevant computational methods applied to 3-D metamaterial design. Table~\ref{table_numerical} summarizes the main features of these methods.  In~terms of generality of use, robustness, and accuracy,  finite difference, finite element, and integral-equation methods are the most applicable, normally at the cost of higher computational resources and lack of physical insight. Modal techniques and reduced-order models (circuits, ray tracing, and homogenization) provide faster computation times and physical insight on electromagnetic phenomena, typically in exchange for lower flexibility and accuracy in the computation. These computational methods have been efficiently applied in a wide variety of periodic and truncated 3-D configurations, such as nanoparticle systems, EBG structures, wire media, cavities, chiral elements, and cloaking~devices. 

Second, we presented an overview of the most outstanding designs and inherent functionalities of 3-D metamaterials. A~wide range of applications can be covered, from~classical filters to novel metasurfaces for polarization control. Three-dimensional architectures have been  theorized for several decades, but~the \emph{massive} experimental exploitation comes from the beginning of the current century. In~this sense, they originally played an important role in the discovery and development of metamaterials thanks to the exotic electrical properties of 3-D distributions of split-ring resonators. The~existing advantages with respect to classical 2-D designs makes 3-D structures potential candidates for operation in future applications, especially in spacial environments and beyond millimeter frequencies, where dielectric materials are undesired. The~work realized and published during the last decade has demonstrated that low-profile 3-D metamaterials/metasurfaces may already cover a wide variety of functionalities. The~development of novel 3-D devices and prototypes is therefore expected to increase and to be refined due to the permanent improvement in 3D-printing techniques, leading to a new scenario in the fields of~telecommunications.    

Finally, we discussed on the current fabrication techniques for the prototyping of 3-D metamaterials. A~wide variety of manufacturing techniques have been used in the work reported in the literature. Of~all those found, the~most relevant are PCB fabrication, which requires a mandatory three-dimensional assembly of the fabricated layers, and~3-D printing, which by its nature fits the 3-D geometries that appear in the unit cells of the metamaterials. Regarding the latter, 3-D printing in plastic with a subsequent metallization is the manufacturing strategy that predominates in the reviewed works, essentially due to the reduced cost compared with direct 3-D printing on~metal.

\vspace{6pt} 



\authorcontributions{Conceptualization, A.A.-A. and C.M.; computational methods (Section \ref{sec2}), A.A.-A.; designs (Section \ref{sec3}), C.M.; fabrication techniques (Section \ref{sec4}), Á.P.-C.; introduction and future trends (Sections \ref{sec1} and \ref{sec5}), A.A.-A., Á.P.-C., and~C.M. All authors reviewed and edited the~document. All authors have read and agreed to the published version of the manuscript.}

\funding{This work was funded in part by the Predoctoral Grant FPU18/01965 and in part by the financial support of BBVA Foundation through a project belonging to the 2021 Leonardo Grants for Researchers and Cultural Creators, BBVA Foundation. The BBVA Foundation accepts no responsibility for the opinions, statements, and contents included in the project and/or the results thereof, which are entirely the responsibility of the authors.} 

\institutionalreview{Not applicable}

\informedconsent{Not applicable}


\dataavailability{Not applicable} 

\conflictsofinterest{The authors declare no conflicts of~interest.} 


\begin{adjustwidth}{-\extralength}{0cm}
\reftitle{References}

\end{adjustwidth}
\end{document}